\documentstyle[aps,epsf,epsfig]{revtex}
\begin{document}
\draft
\twocolumn


\title{Stripes in Quantum Hall Double Layer Systems}

\author{L. Brey}
\address{Instituto de Ciencia de Materiales de Madrid, 
CSIC,
28049 Cantoblanco, Madrid, Spain.}

\author{H.A. Fertig}
\address{Department of Physics and Astronomy,
University of Kentucky, Lexington, KY 40506-0055}

\date{\today}

\maketitle

\begin{abstract}
\baselineskip=2.5ex

\begin{abstract}

We present results of a study of 
double layer quantum Hall systems in which 
each layer has a high-index Landau level that
is half-filled.  Hartree-Fock calculations
indicate that, above a critical layer
separation, the system becomes unstable to the
formation of a unidirectional coherent charge density
wave (UCCDW), which is related to
stripe states in single layer systems.
The UCCDW state supports a quantized
Hall effect when there is tunneling between
layers, and is {\it always} stable
against formation of an isotropic Wigner
crystal for Landau indices $N \ge 1$.
The state does become unstable to the formation
of modulations within the stripes at
large enough layer separation.  
The UCCDW state supports
low-energy modes associated
with interlayer coherence.  The coherence
allows the formation of charged soliton
excitations, which become gapless in
the limit of vanishing tunneling.  
We argue that this may result in a
novel {\it ``critical Hall state''},
characterized by a power
law $I-V$ in tunneling experiments.

\end{abstract}

\vspace{0.5truecm}
\end{abstract}
PACS number  73.40.Hm, 73.20Dx, 73.20Mf
]
\section{Introduction}
Recently, it has been established that quantum Hall systems in
which several Landau levels are filled may support states
with highly anisotropic transport properties\cite{expstr}.
High mobility two-dimensional
electron systems, in
perpendicular magnetic fields such that the filling
factor $\nu$ is in the vicinity of $N+1/2$,
with $N$ an odd integer greater than 7 \cite{tilt},
are found to have diagonal resistivity ratios $\rho_{xx}/\rho_{yy}$
as large as 3500 \cite{lilly2}.  ($\nu \equiv N_e/N_{\phi}$ 
is the ratio of the number of electrons to the orbital 
degeneracy of a Landau level $N_{\phi}=AB/\Phi _0$, 
where $B$ is the magnetic field, $A$ is the sample area
and $\Phi _0$ is the magnetic flux quantum.)
While a good deal of this anisotropy is purely
geometric\cite{simon}, the experiments clearly
indicate a groundstate for the electrons of a
highly anisotropic nature.

These results are very likely to be related to stripe phases
that have been found in mean-field 
calculations \cite{shklovskii,moessner}
and exact diagonalization studies \cite{rezayi} of electrons
with partially filled high Landau levels.
When thermal and quantum fluctuations are included,
it is useful to classify possible states of such systems
according to their symmetries, which bear a close
resemblance to those found in liquid crystals \cite{fradkin}.
Of particular interest is the smectic state, which,
in the presence of some (hitherto unidentified)
orientational ordering field, should exhibit strong
macroscopic transport anisotropies as observed
in experiment.
Quantum fluctuations in this state have been
studied, with some
models \cite{fertig,emery} indicating the 
possibility of metallic behavior in the groundstate.
(See, however, Ref. \cite{AHM1}.)  The mean-field
stripe states are a natural first approximation
to such smectic states.

It by now well appreciated that for smaller filling factors,
the introduction of a second layer in the quantum Hall
system (i.e., double quantum well systems) 
leads to new phenomena and a rich phase 
diagram \cite{tlrev}.  
For a partially filled lowest Landau level ($N=0$),
at total filling factor $\nu=1$ and for small layer
separation $d$, a spontaneously broken symmetry
can occur in which interlayer phase
coherence develops in the absence of interlayer
tunneling.  In this state
the system supports a Goldstone mode,
which has a roton minimum at a 
wavevector $q \approx 1.2\ell$. 
For separations $d>d_c \sim \ell$, this minimum goes to
zero, indicating a phase transition  \cite{theory1} into a 
unidirectional, (interwell phase) coherent
charge density wave (UCCDW). 
At still higher
separations, there is another phase transition
into a phase-coherent Wigner crystal 
state (CWC) with two non-collinear primitive
lattice vectors\cite{chen1,cote1}.

For $N$=0, these states with broken translational symmetry are 
usually preempted by
strongly correlated Fermi-liquid states in which translational invariance is 
restored by quantum fluctuations\cite{hlr}.
However, in higher Landau levels the effective 
interaction among electrons in a single layer
contains structure that favors such a
mean field state 
relative to uniform correlated 
states\cite{fogler,jung}.
This makes the double layer electron system in
higher Landau levels an excellent
candidate for observation of 
the interesting translational broken-symmetry 
states described above.

In this article, we
demonstrate within the Hartree-Fock approximation that
such states do indeed occur for electrons systems
with high filling factors.  
We will focus on systems
with $\nu$=$2 \nu _L$=$2\times(2N+1/2)$, 
where $N$=0,1,2,...
$N$ is the index of the highest occupied orbital
Landau level (LL) and $\nu_L$ is the filling factor in each layer.
Unlike the situation
for single layers, we find  
circumstances under which the system supports
stripe ordering {\it and} a quantized Hall
conductance simultaneously.  Physically, this 
arises due to the development of coherence between 
the edges
of the individual stripes, which opens up a gap
in the single particle spectrum.  This coherence
has interesting consequences for the collective
modes of the system, as well as for its
low-energy charged excitations.

The results of this work may be summarized as follows:
\par \noindent
i) For any value of $N$, and for small layer separation, 
the groundstate of this system has a uniform density and
phase coherence between the layers.  The system will
support a quantized Hall conductance.
\par \noindent
ii) In the limit where tunneling may be neglected,
the energy spectrum of the system contains a gapless
linear (Goldstone) mode due to the spontaneous interlayer
coherence.  This mode contains a roton minimum at wavevectors
$q \sim 1/\ell$, which vanishes when the layer separation
$d$ exceeds some critical value $d_c$.  The precise value
of $d_c$ depends on $N$.
\par\noindent
iii) For $d>d_c$, a state with unidirectional charge density wave
order {\it and} phase coherence between the layers is lower in energy
than the uniform density state.  For $N \ge 1$ and any value 
of the layer separation, the coherent unidirectional
charge density wave has lower energy than the square or triangular
Wigner crystal.  For tunneling parameter
$t \ne 0$, this state supports
a quantized Hall effect.
\par \noindent
iv) For small values of $t$, the 
linear regions in which the density has significant magnitudes
in {\it both} layers -- i.e., the regions between the centers
of stripes in different layers -- will support low energy collective
modes.  These excitations are 
analogous to spin
waves of one-dimensional $XY$ magnets in a magnetic field.
They sustain soliton excitations
which carry charge $\pm e$, whose energy scales as $\sqrt{t}$.
\par\noindent
v) For $t=0$, the charged excitations become gapless
and no quantized Hall effect can occur.  
At zero temperature the system may undergo a Kosterlitz-Thouless
transition, in which, in the equivalent 1+1 dimensional
{\it classical} {XY} system \cite{sondhi}, vortices
unbind.  The bound vortex state is very unusual,
similar to quantum Hall states in having
a vanishing tunneling conductance, but nevertheless
being gapless and not supporting a plateau in the
Hall conductivity.  Because of the power law correlations
inherent in systems with bound vortices, we call this
a {\it critical Hall state}.
\par\noindent
vi) At higher values of the layer separation the 
electron system undergoes a
first order phase transition to a non-coherent
unidirectional charge-density wave state with a weak 
longitudinal modulation
of the charge density. This state also can be 
viewed as a highly anisotropic
Wigner crystal (i.e., a stripe crystal.)

This paper is organized as follows.  In Section II we describe the 
uniform coherent state and study its charge 
density excitations.
~From the softening of this excitation we 
obtain a phase diagram, in tunneling versus 
layer separation, for which the UCS 
becomes unstable. 
In Section III we discuss in 
more detail the method of 
calculation used for studying the
translational broken symmetry ground 
states, and introduce the different
phases analyzed.
In Section IV we present our numerical 
results.
Section V is devoted to a qualitative
discussion of the excitations
supported by the unidirectional coherent
charge density wave state, and their 
consequences for quantum fluctuations.
We conclude in Section VI with 
a summary.

\section{Instabilities of the Uniform Coherent State.}

In what follows, we consider our double layer system at total
filling factor $\nu=4N+1$.  The $4N$ part of this filling refers
to the fully filled Landau levels, each of which has two spin
states and two ``layer states''.  
In the large magnetic field limit, it is safe to assume these
lower levels are filled and inert, and we
do not explicitly include them in the
calculations below.  We also
assume there is no
spin texture \cite{skyrmion} in the groundstate,
which is always true if the Zeeman coupling
is large.  If the tunneling amplitude $t$ is
smaller than
the Zeeman coupling, then our results  
apply to filling factors $\nu=4N+3$ {\it mutatis mutandis}.

For small separations between the layers, the 
groundstate of the system is expected
to be a  uniform coherent state, with wave function of the form 
\begin{equation}
|\Psi _N > = \prod _ X   { {1 \over {\sqrt 2}} 
\left ( C _ {N,X,l,\uparrow} ^ +  \, + \, 
e ^{ i \theta} C _ {N,X,r, \uparrow} ^+  \right     )}  |N-1> \, \, \, 
\end{equation}
where
\begin{equation}
|N-1>\,  = \prod _{ X';\, \sigma=\uparrow \downarrow ;\, N'<N ;\, i=l,r}  
C _ {N',X',i,\sigma} ^ + |0> \, \, \,.
\end{equation}
In these equations $ C _ {N,X,l(r),\sigma} ^ +$  creates an  
electron in the Landau level (LL) $N$, with 
guiding center orbital label $X$ and spin $\sigma$ 
in the left (right) well.
$|0>$ is the electron vacuum and $|N-1>$ is the
wavefunction which describes
the system at filling factor $4N$, for which 
all the states with LL orbital index $N'<N$ are occupied.
Note that since in Eq.(1) the product runs over all the 
possible values of $X$ this state corresponds to filling 
factor $\nu=4N+1$.
By construction this state has the same electronic charge 
in each well.
Since
we assume that the only active electron states are those 
in the highest LL 
with spin parallel to the field, 
in order to simplify the notation
we drop the indices $N$ and $\uparrow$ 
in the creation and destruction 
operators in what follows.

The state described by Eq.(1) has long-range order in the phase difference
between electrons in the two layers, $\theta$. 
In the absence of tunneling between layers,  
the energy is independent of $\theta$, and the
system has a continuous broken symmetry. 
For $t \neq 0$, 
the ground state of the system is composed of a full LL of
single-particle orbitals which are 
a symmetric combination of the left and
right layer orbitals ($\theta=0$). 
This wavefunction is exact for non-interacting electrons,
and for $d \rightarrow 0$ we expect it to remain exact 
when interactions are included.
For finite $d$, quantum fluctuations become important and 
the ground state
described by Eq.(1) is only
an approximation. Nevertheless,
we expect \cite{theory1} a 
broken symmetry ground state  
to survive for $d$ 
smaller that a critical value of the 
layer separation $d_c$.

In order to estimate $d_c$ we calculate the 
charge density excitations (CDE's) 
of the system.
The CDE's are classified by a conserved wavevector 
${\bf q}$ \cite{kallin},
and their 
dispersion can be obtained from the poles of the 
charge density response function.
Neglecting mixing between LL's,  it has the form
\begin{equation}
\omega _{cde} (q)\! = \!
\left \{ [ \Delta ^{HF} \!- \!V_d (q) ]
[\Delta ^{HF} \!-\! V _b(q)\! +\! V_a(q) \!-\! V_c(q) ] \right \} ^ {1/2} 
 ,
\end{equation}
where
\begin{eqnarray}
\Delta  ^{HF} & = &  2 t   + V_d (q=0) \, \, \, , \nonumber 
\\
V_a(q)&  = & {{ e ^2} \over {\epsilon \ell}}  { 1 \over {q \ell}} V(q) \, \, \, , \nonumber
\\
V_b(q) &= &{{ e ^ 2} \over {\epsilon \ell}}  \int _0  ^ {\infty} d (q'\ell) 
J_0 (qq'\ell^2) V(q') \, \, \, , \nonumber
\\
V_c (q) &= &{{ e ^2 } \over {\epsilon \ell}}  { 1 \over {q \ell}} V(q) e ^{ -q d} 
\, \, \, \nonumber 
\\
V_d(q )&  = & {{ e ^2 } \over {\epsilon \ell}}  \int _0 ^{\infty} d (q'\ell)J_0 (qq'\ell^2) V(q') e ^{ -q' d} \, \, \, \nonumber
\end{eqnarray} 
with
\begin{equation}
V(q)= e ^{ - q^2 \ell ^2 /2} \left  ( L_N ( q ^2 \ell ^2 /2) \right  ) ^ 2 
\, \, \, ,
\end{equation}
where $L_N(x)$ is a Laguerre polynomial.
In these expressions $\ell =\sqrt { { \hbar c / e B}}$ is the magnetic length
and $\epsilon$ is the dielectric constant of the host semiconductor.
$V_a$ and $V_b$ are the direct and exchange intrawell Coulomb interaction,
while $V_c$ and $V_d$ are the direct and exchange interwell 
Coulomb interactions \cite{cote1}.

We have computed $\omega _{cde}(q)$ for different values of $N$,
and have plotted several representative curves 
(with $t=0$) in Fig.1 for different 
values of 
$N$ and for a layer separation near $d_c$. 
For $t=0$,  at small
wavevector $q$ and any value of $N$, the dispersion curves
increase linearly. This acoustic excitation is the  Goldstone
mode associated with the continuous broken symmetry;
i.e., the spontaneous interlayer coherence. 
At intermediate $q$ the dispersion curves develop 
a dip which becomes 
soft at a wavevector $q_c$ 
for values of $d$ bigger than a 
critical
distance $d_c(N)$.
As discussed above, this indicates
an instability of the state.
We can see in Fig.1 that, in 
magnetic length units, the value of $d_c$ decreases 
as the LL index increases.
At larger values of $q$ (not shown in Fig.1) the spectra
corresponding to $N > 0$ develop some structure 
associated with the zeros of the 
Laguerre polynomials appearing in the definition of
$V(q)$.

Fig. 2 illustrates $d_c$ versus 
$t$ for a few values of $N$, and thus represents
a phase diagram for
the stability of the uniform coherent state
(stable for small $d$ and large $t$)
with respect to the formation
of a density wave state (for large $d$ and small $t$).
Note that the uniform state is stable
for $N=2$ even at $t=0$.  This is {\it not} the
case for single layers, where at $\nu=1/2$
the system forms stripes.  Evidently,
the condensation energy for spontaneous
interlayer coherence outweighs the gain in energy
associated with stripe formation in the
individual layers.  For large
enough $d$, however, the single layer
tendency to form stripes must eventually
become important.
In order to 
more precisely
characterize the ground state 
after the instability sets in,
we study different 
broken translational symmetry states.
This is the subject of the next Section.

\section {Broken Translational Symmetry States}

In this Section we study different translational broken symmetry states
of the electron gas confined in a DQW system.
We adopt the Hartree-Fock approximation in the form 
introduced by C\^ot\'e and
MacDonald \cite{cote2}; in this method the
important quantities are
the expectation values 
of the operators
\begin{equation}
\rho _{j,j'} ({\bf q}) = { 1 \over {N_{\phi}}}
\sum _{X,X'} e  ^{- i q_x (X+X')/2 } 
\delta _{X,X'-q_y \ell ^2} C^+ _{X,j} C_{X',j'} \, \, \, .
\end{equation}
Here the quantum numbers $j,~j^{\prime}$ are layer indices.
Details of the application of this method
to bilayer systems may be found in Ref. \cite{cote1}.


The Hartree-Fock equations have a number 
of different solutions corresponding
to different states of the electron gas 
in the DQW system. Each 
solution is characterized by a set of order parameters
$< \rho _{i,j} ({\bf q })>$; the ground state energy 
can be expressed
solely in terms of these quantities. In the 
following calculations we consider a limited
number of physically interesting solutions and compare 
their energies, and consider situations 
when transitions among these different
states may occur.

\subsection{Uniform Coherent State (UCS).}

This state is 
described by Eq.(1); it is characterized 
for $\theta=0$ by
$<\rho_{l,l}(0)>=<\rho_{r,r}(0)>=<\rho_{r,l}(0)>= \, \, 
<\rho_{l,r}(0)>=1/2$, and all other order 
parameters zero. As discussed above,
translational invariance is not 
broken in this phase, but there is interwell
coherence which, at $t=0$, breaks a $U(1)$ symmetry 
of the Hamiltonian.  In Eq. 1, this arises
because the energy of the state is
independent of $\theta$ \cite{theory1}.
The UCS has always lower energy 
than the incoherent, $<\rho_{l,r}(0)>=0$, 
uniform state.

\subsection{Unidirectional Coherent CDW State (UCCDW)}
In this state
$<\rho_{i,j} (n {\bf G}_0) > \neq 0$ , $n=0, \pm 1, \pm 2, ...$; 
the translational symmetry in one direction is broken 
{\it and} interwell coherence is 
allowed. The value of $G_0$ is chosen 
to minimize the energy of the system.
Furthermore, in order to have an energy
minimum in this class of states,
the CDW in the two wells must be shifted
by a distance $\pi/G_0$ with respect to one another.
For $d \rightarrow d_c$ the value of
$G_0$ which minimizes the energy approaches $q_c$.

The UCCDW state may be written in the form
\begin{equation}
|UCCDW\!>\! =\! \prod _C\! \left ( \!\sqrt {\nu _l (X)} c ^+ _{X,l} \! 
+ \! \sqrt {\nu _r (X)} e^{i\theta} c ^+ _{X,r}
\right )\! |N \!- \!1\! > 
\, \, \, \, , 
\end{equation}
where $\nu _{l(r)} (X)$ is the occupation of the state 
$X$ in the left (right) well,
and $\nu _l (X)+\nu _r (X)=1$. The local occupation factors
$\nu _i (X)$ are periodic functions of $X$ with period $2\pi /G_0$. 
With an appropriate choice of origin,
the charge density is an even function of $X$, 
so the state may be constructed with
only odd Fourier components of 
$< \rho _{i,i} (nG_0)>$. The local coherence between wells is 
given by the quantity 
$\sqrt {\nu _l (X) \, \nu _r (X) } e^{i\theta}$, which is
even in $X$ and periodic with period  $\pi/G_0$;
thus 
$<\rho _{l,r} (n G_0)> \ne 0$ only when $n$ is even.

~From the order parameters it is easy to 
obtain the band structure of the UCCDW phase.
The eigenvalues depend on the guiding center
coordinate $X$, and 
(up to an unimportant constant) are given by 
\begin{equation}
\epsilon (X) = \pm \sqrt{ { A(X)} ^ 2 + { B (X)} ^ 2} \, \, ,
\end{equation}
where
\begin{eqnarray}
A(X) &\! =\! &  \sum _{n \ne 0} e ^{ i n G_0 X} \! \left [
V_a (nG_0)\! -\! V_b (nG_0)\!-\! V_c (nG_0) \right ] \!
<\!\rho _{i,i} (n G_0)\!> \, \, ,
\nonumber  \\
B(X) &\! = \! & -t - \sum _n V_d (nG_0) < \rho _{l,r} (nG_0)> \, \, \, \, .
\nonumber
\end{eqnarray}
Since the total filling factor of the $Nth$ Landau
level is 1, only the lowest energy band
is occupied.
In the UCCDW phase we define the gap, $\Delta _{UCCDW}$,  as the difference 
in energy between
the highest occupied state and the lowest empty state.
In the coherent solution, $<\rho _{l,r} (0)>$ is always different from zero, 
so that $B(X) \ne 0$ for all $X$;
in this case $\Delta _{UCCDW} > 0$.
It is important to recognize that the 
Hartree-Fock approximation overestimates
the actual gap to create well-separated particle-hole
pairs above the UCCDW state, in particular because
the groundstate band structure does not reflect the 
possible existence of soliton quasiparticles.  As 
we discuss below,
for $t>0$, a magnetic field-dependent gap is
always present and a quantized Hall conductance will
be observed \cite{AHM}; for $t=0$ we shall
that the quantized 
Hall effect is absent. 

In the limit $d \rightarrow \infty$,
$\nu _l (X) \nu _r (X)=0$ and the coherence between wells
is lost;
$\nu_i(X)$ can only be zero or one. This limit corresponds to two uncoupled
2DEG's each supporting an independent stripe state.

\subsection{Coherent WC state (CWC)}

In this state 
$\{<\rho_{i,j} ({\bf G}) > \} \neq 0$,  
translational symmetry is 
broken along two non-collinear vectors, 
and interwell coherence is allowed.
The reciprocal lattice vectors $\{ {\bf G} \}$ 
define a two-dimensional lattice 
containing one electron per unit cell. 

At intermediate values of $d$ the square lattice 
has lower energy than the triangular lattice,
because the interstitial regions in a square lattice
are larger than those of a triangular lattice,
allowing a particularly low interwell Hartree
energy \cite{dwwc}. 
At larger values of $d$ 
there will necessarily be a first order
phase transition into a triangular lattice,
since its intralayer Madelung energy is
lower than that of the square lattice.
However, because the Madelung energies of
the two states are very close \cite{bonsall},
this only occurs at a very large value of $d$.
We thus do not further consider
the triangular lattice for the
purposes of this work; 
the Wigner crystals we discuss
are square lattices,
with unit cell lattice parameters $a_x=a_y=2 \sqrt {\pi} \ell$. 

As in the case of the UCCDW, in order to minimize the 
interwell electrostatic 
energy, the lattice in the two wells are shifted
by a vector $(a_x/2,a_y/2)$.

For systems modulated in two directions, 
such  as the CWC phase, there are gaps
in the excitation spectrum, but the densities at 
which they
occur are not $B$ dependent and the Hall 
conductance is not expected to be quantized.

\subsection{Modulated Unidirectional CDW state. (MUCDW)}

In this state the electron gas has modulations 
in the charge density
along the direction of the stripes. 
Such modulations are known to
significantly lower the energy
of stripes in a single layer \cite{cote_modes}.
While the two-dimensional long-range order that appears
in such a state may not survive in the
groundstate due to quantum
fluctuations \cite{fertig,emery,AHM1},
the appearance of such modulations
within Hartree-Fock reflects the tendency of 
the stripes to look like one-dimensional
crystals at short length scales.

In the calculation we use 
either
an oblique 
or a 
rectangular lattice with one electron per unit cell.
In either case, 
the primitive lattice vectors have the form
\begin{eqnarray}
{\bf a }_1  & = & (X_0, Z_0 ) \nonumber 
\\
{\bf a }_2  & = & (0,  Y_0  ) \, \, \, \, , \nonumber 
\end{eqnarray}
with $X_0 = 2 \pi /G_0$ and $Y_0= 4 \pi \ell ^2 / X_0$.
The parameter $Z_0=Y_0/2$ for an oblique lattice, 
and $Z_0=0$ for a rectangular lattice.
In single layer systems, it is found \cite{cote_modes} 
that the energy is extremely insensitive to the
parameter $Z_0$, but there is a very shallow
minimum for $Z_0=Y_0/2$.  This turns out to
be the case of the present system as well.

For the two layer system
under consideration here,
to minimize the interwell Madelung 
energy the charge in the two wells is shifted
with respect to one another by a vector
\begin{equation}
({ { a _1 ^ 2 } \over {2 X_0}}, 0) \, \, \, \, . \nonumber
\end{equation}
This particular shift vector is the analog of
what one finds in hexagonal close packed 
structures.  

In this class of solutions we have never found a coherent state.
Whenever $<\rho _{i,i} (G_x,G_y) > \ne 0$ for  $G_y \ne 0$, we always find
$< \rho _{l,r} ({\bf G})>  =0$ for all set of ${\bf G}'s$.
Physically, this arises because formation of modulations
competes with
interwell coherence.  The latter lowers the energy
by introducing admixtures
of single particle states near the Fermi surface in
different wells.  Modulations, by contrast, {\it destroy}
the Fermi surface: modulated stripes essentially represent a
highly anisotropic Wigner crystal state, with gaps everywhere
in the band structure between occupied and unoccupied
states.  Thus, there are no low-energy states that may
be admixed between wells.
The MUCDW is a solution without coherence between the wells and 
we find that 
it is not possible to change continuously from the UCCDW phase to the MUCDW phase.

As in the CWC phase, the gap in the excitation spectrum which 
appears in the MUCDW phase
is present over a range of densities.  
This phase does not exhibit a quantized Hall effect.

\section{Results.}

We now present our numerical results for the states
discussed in the last Section.
In Fig.3 we plot the energy difference
$E_{CWC}-E_{UCCDW}$ as a function of the layer separation for different
values of $N$. A negative (positive) value 
of this quantity implies that the CWC has 
lower (higher) energy than the UCCDW state.
Because the 
transition at $d_c$ is continuous,
both energies, $E_{CWC}$ and $E_{UCCDW}$, tend 
to the energy of the UCS at $d=d_c$.
For any value $d$ near but higher than $d_c(N)$, 
we find the UCCDW phase has
lower energy than the CWC, indicating 
the instability at $d_c$ is  
a  transition
from the UCS to the UCCDW state. 
The driving energy for this transition 
comes from the difference between intra- and inter-well interactions,
which increases with layer separation. For small $d$, interwell and intrawell
interactions are rather similar; at larger layer separations lower
energies can be attained by
improving intrawell correlations via 
$<\rho_{i,i} (G_0)>$ taking on a non-zero value.
This can be accomplished only by allowing the phase relationship
between electrons in different wells to fluctuate, i.e., by  lowering
the expectation value of $<\rho _{l,r} (0)>$ (cf. Eq. 7).

~For $N\!=\!0$ a second transition occurs 
from the UCCDW state to the CWC state
at $d \sim 1.65 \ell$. As commented above  
in the $N\!=\!0$ case the fluctuations will
probably melt this CWC state, forming two non-coherent  
highly correlated 
Fermi liquids \cite{hlr}.
For $N\ge 1$, the situation is different: there is no  
UCCDW$\rightarrow$CWC
transition. We find that the UCCDW phase always has 
lower energy than the CWC state.
The difference in energy between these states increases with $N$. 
Obviously this is closely related to the fact that 
for $N\ge 1$, in the HF approximation 
the stripe phase is lower in energy 
than the WC state \cite{shklovskii,moessner}.

We now discuss the properties of the UCCDW state
in more detail.  For concreteness, 
we focus on the case
$N=2$, $d=1.2 \ell$ and $t=0$.  Most 
of the qualitative results, however, apply 
to all the UCCDW states for other
parameters.

Fig. 4 illustrates the local filling factor in the
left well $\nu _l (X)$ and the local coherence factor
$\sqrt{ \nu _l (X) \nu _r (X)}$ as a function of $X$. 
For this case $G_0 \sim  \ell ^{-1}$, and the 
width of a unit cell is
$ \sim 2 \pi \ell$. Note that
$\nu _l (X)$ is the particle-hole conjugate 
of $\nu_r(X)$, i.e., $\nu_r (X)=1-\nu_l (X)$.
The coherence between the wells is maximum 
in the region where the 
charge density is shared by both wells. Note that, 
as commented above,  the interwell coherence order parameter has
periodic $\pi /G_0$. For smaller values of $d$, the oscillations of the density
between wells decreases until, at $d_c$, the charge density 
becomes uniform, with
$\nu_l (X)=\nu_r(X)=1/2$.  At this point
the coherence factor becomes constant 
and takes on its maximum value of $1/2$.
For larger values of the layer separation, $\nu_{l(r)} (X)$ 
changes more rapidly, and the
coherence is significantly different from zero 
in thinner regions.  As 
$d \rightarrow \infty$, the 
width of the coherent regions tends
to zero, and the local occupation is
just zero or unity.

In Fig. 5 we plot the band structure, Eq.(9), in the UCCDW case.
For the parameters used in the calculations the
Hartree-Fock results show
a gap for the charged excitations.
The gap is maximum in the regions 
where the charge is localized in one of the wells
and minimum in the regions where the charge is shared 
by both wells. For small values of $d$, 
the band structure becomes flatter and 
the energy gap increases until, at $d_c$,
the gap is constant in the Brillouin zone 
and takes its maximum value $2(t+V_d(0))$.
For larger values of $d$ the gap decreases,
reaching its minimum value $2t$ 
when $d \rightarrow \infty$. In the UCCDW phase, 
$<\rho_{l,r}(0)> \neq 0 $, so that  there is 
always a finite energy gap, although in some circumstances
this gap is much smaller than the typical experimental temperatures.
It should be noted that the spatial structure
of this gap suggests that charged quasiparticles,
either due to thermal fluctuations or to doping
away from total filling factor $\nu=4N+1$, will
be much more mobile along the direction of the
stripes than perpendicular to them.  It would
be most interesting if such anisotropy in
double well transport could be observed
experimentally.

It should be noted that in Eq. 6, the wavefunction
for the UCCDW has a free parameter $\theta$, whose
value for $t=0$ does not affect the energy.  This
implies that the UCCDW supports gapless modes
(beyond phonons), due to the spontaneous coherence.
The interesting aspect of this system is that,
for large enough $d$, the coupling between the
coherent regions will become negligible, and
each linear region may support its own
gapless mode.  As discussed below, the resulting system
is similar to a collection of one-dimensional $XY$ ferromagnets.
The fluctuations of this degree of freedom greatly
reduces the charge gap for any value of $t$ compared
to what is found in the band structure, and for
$t=0$ we predict that such fluctuations will drive
the gap to zero.  

Returning to the Hartree-Fock results,
the physical origin for the existence of a CDW in DQW 
systems is easy to understand. For  $d$ near $d_c$
the Hartree-Fock one body potential has a minimum at a 
wavevector smaller than the reciprocal lattice
vector associated with a WC, and the electron gas  
can attain a lower energy
by forming a UCCDW. 
As $d$ increases, the minimum in the effective potential
moves to larger wavevectors. 
Because of this, for  $N=0$ the WC becomes 
more stable than the unidirectional CDW.
For $N\ge1$, by contrast, 
the effective two-dimensional Coulomb 
interaction, $V(q)$, contains Laguerre polynomials 
that have zeros at decreasing 
wavevectors when $N$ increases. This 
produces a zero in the repulsive Hartree 
potential where the exchange interaction is strong,
making the UCDW more stable than the WC
at larger layer separations.

In Fig.6 we plot the order parameter $<\rho _{l,r} (0)>$ and the energy gap 
$\Delta _{UCCDW}$ as a function of the layer separation. Note that in the
UCCDW phase the coherence is non-zero even for large values of $d$.
As $d$ increases the charge energy gap goes to its minimum value, 
but this is not so much due to the decrease in the interwell coherence 
as it is to
the increase of the intrawell coherence $<\rho _{i,i} (G_0) >$.

Finally, we address the
question of the value of the layer separation for which
the two electron layers become decoupled, forming at $N>0$, a pair
of noncoherent MUCDW states. To find this, one must compute
the energies of the MUCDW and the UCCDW as 
functions of the layer separation. 
These may be found in Figs. 7 and 8  
for the cases $N=2$ and $N=1$,
respectively. 
As  commented above, as a function of $d$,  
there is a crossing (first order phase transition) 
in energies between the MUCCDW and the UCCDW phases.  
The transitions occurs at $d \sim  1.6 \ell$ in the $N=2$ case and at
$d \sim 1.4 \ell$ in the $N=1$ case, indicating that the 
UCDW phase becomes more
stable as the LL index increases. 
~From the value of the charged excitation gap, we expect that 
the QHE in 
double layer systems for $t \ne 0$
will disappear for large enough $d$
due to the UCCDW$\rightarrow$MUCDW transition and not
the UCS$\rightarrow$MUCDW transition.

\section{Collective Modes and Solitons in the UCCDW State}

As mentioned above, the UCCDW state contains an interesting
degree of freedom, the phase of the coherence factor in
the regions where there is significant occupation of
both the left and right wells (cf. Fig. 4).  As in the
case of the UCS, this phase may be thought of as a function
of position, and from it low energy excitations may
be constructed \cite{tlrev}.  Interesting new physics
emerges from this degree of freedom in the limit of
large $d$, when the linear coherent regions become very
narrow, and the magnitude of the coherence factor
nearly vanished between these regions.  In this situation
we may view the coherent regions as a collection of
one-dimensional systems which are, in a first approximation,
uncoupled \cite{com}.  Each
coherent region thus has its own phase, $\theta_i(y)$, where
$i$ labels the individual coherent region, and we assume the
stripes are parallel to the $\hat{y}$ direction.

Once we adopt a model in which the $\theta_i$'s are uncoupled,
the low energy statics and dynamics become identical to
those of a domain wall between spins of opposite
orientation in a filled Landau level.  This system
has been studied in Ref. \cite{falko}, and we may
use those results to understand what will happen in 
the double layer system.  At long wavelengths, the
energy functional for the phase takes the form
\begin{equation}
E_i = \int d y \Biggl[ {1 \over 2} \rho_s \Bigl( {{\partial \theta_i} \over
{\partial y}} \Bigr)^2 - g t \cos{\theta_i} \Biggr].
\label{e_i}
\end{equation}
In the above equation, $\rho_s$ is a stiffness
for the effective $XY$ spin degree of freedom, 
and $g$ is analogous to a Land\'e $g$ factor, with $t$ playing
the role of a magnetic field.  The
values of both $\rho_s$ and $g$ depend on the 
precise form the UCCDW wavefunction, and require
a microscopic calculation to evaluate.  Note both
$\rho_s$ and $g$ remain finite in the limit $t \rightarrow 0$.

Eq. \ref{e_i} has the form of a sine-Gordon
model, and a number of interesting properties immediately
follow.  Of special importance is the fact that
the system supports soliton excitations, where $\theta_i$
changes by $2\pi$ in $y$ going from $-\infty$ to $\infty$
\cite{raja}.  The energy of these excitations have the
form $\varepsilon_s=8\sqrt{\rho_s gt}$.  For reasons
very similar to what occurs in double
layer quantum Hall systems near $\nu=1$ \cite{tlrev},
or analogously for systems with spin \cite{skyrmion2},
the sine-Gordon model solitons carry a net charge
of $\pm e$ \cite{falko}.  This means the linear coherent
regions support charged excitations {\it much} lower
in energy than predicted by the Hartree-Fock theory
for small $t$: as discussed above, the energy to create
a well-separated particle-hole pair tends to
a constant as $t \rightarrow 0$, while the soliton
energy vanishes as $\sqrt{t}$.  For small $t$, at
finite temperature the solitons should dominate the
transport properties of the system.

In the limit of vanishing tunneling amplitude, a very
interesting possibility arises for behavior of
the linear coherent regions.
As stated above, the system will have gapless modes associated
with the spontaneously broken symmetry; these lead \cite{falko}
to a set of collective modes dispersing linearly:
$\omega(q) = A q$ \cite{com2}.  The action consistent
with this has the form 
\begin{equation}
S=\int d\tau dy 
\Biggl[ {1 \over 2} \rho_{\tau} \Bigl( {{\partial \theta_i} \over
{\partial \tau}} \Bigr)^2 +
{1 \over 2} \rho_s \Bigl( {{\partial \theta_i} \over
{\partial y}} \Bigr)^2 \Biggr],
\label{action}
\end{equation}
where $\tau$ is imaginary time, and $\rho_{\tau}$ is an
effective moment of inertia that depends on the
precise form of the UCCDW wavefunction.  
At zero temperature, Eq. \ref{action}
may be reinterpreted as an energy functional in 
a 1+1 dimensional {\it classical} system \cite{sondhi}.
This system will undergo a Kosterlitz-Thouless transition
if $\sqrt{\rho_s \rho_{\tau}}$ passes through $2/\pi$.
Above this value, vortices in the $\theta_i$ field
are bound in pairs; below it, they are unbound.

The meaning of this phase transition may be understood
by examining what happens for small but non-vanishing
$t$.  In this case, vortices are bound by a linear
potential, and there will be a ``string'' attaching
them in which $\theta_i$ rotates by $2\pi$ \cite{tlrev}.  
~bFrom the spin-charge coupling inherent in such multicomponent
systems, this means the vortices ``cap'' a region in
spacetime in which the total charge in the $i-th$ linear
coherent region fluctuates away from the average by
$\pm e$.  For $t>0$ the vortices always remain bound,
and these fluctuations are controlled.  Thus, fluctuations
in the charge density of the 2DEG are suppressed, and
the system exhibits a quantized Hall conductance.

For $t=0$ and $\sqrt{\rho_s \rho_{\tau}} > 2/\pi$, the
vortices also remain bound, albeit logarithmically.
This suggests that the charge density fluctuations
are still suppressed, and the system bears 
a resemblance to a quantum Hall phase.  However,
since charged excitations are not gapped at 
$t=0$, the Hall conductance should not exhibit a plateau.
The difference between states with bound and unbound
vortices would be most directly probed by a tunneling
experiment: in the bound-vortex phase, one expects a tunneling
current $I$ that vanishes as a power of the voltage
$V$ that is greater than one, so that the tunneling conductance
vanishes.  
In the unbound state, the presence of free vortices
means charge may be injected into the system, resulting in a finite
tunneling conductance.  Note that for a quantized Hall state,
the tunneling current should vanish as $I \sim e^{-V_0/V}$,
where $V_0$ is proportional to the quasiparticle gap
\cite{he}.  The differing behaviors of the low-voltage
$I-V$ characteristics indicates that the bound vortex
state is qualitatively different than either a conducting
state or a quantum Hall state.  We tentatively call
this new state a {\it critical Hall state}.

In closing this Section, we note that in real systems,
the true groundstate of the DQW can never be a critical
Hall state, because the tunneling amplitude can never
be made to completely vanish in practice.  In addition, there may
be small but relevant gradient couplings among the
different linear coherent regions, driving the system
away from the critical Hall state at low enough temperature.
In practice, both these perturbations may be made very
small, so that at experimentally realizable temperatures
their effects might be unimportant.  It is important to
evaluate whether the system parameters (i.e., 
$t$, $N$ and $d$)
can be adjusted so that properties of a critical
Hall state may be 
observed in practice. This will be the subject
of a future study.

\section{Summary}

In this paper, we studied 
double layer quantum Hall systems in which 
each layer has a high-index Landau level that
is half-filled.  Hartree-Fock calculations
revealed that for small layer separations,
the groundstate is a uniform 
coherent state (UCS).  Above a critical
layer separation $d_c(N)$, we found
a continuous phase transition
to a
unidirectional coherent charge density
wave (UCCDW), which is related to
stripe states in single layer systems.
This UCCDW state supports a quantized
Hall effect when there is tunneling between
layers ($t \ne 0$), and is {\it always} stable
against formation of a coherent Wigner
crystal (CWC) for Landau indices $N \ge 1$.
The state does become unstable to the formation
of a modulated unidirectional charge density
wave (MUCDW) state for
large enough layer separation,
which should lead to the destruction
of the quantum Hall effect. 
 
The UCCDW state supports
interesting low-energy modes associated
with interlayer coherence, which become
gapless in the limit $t \rightarrow 0$.  
For relatively large layer separations,
in a model where the Coulomb interaction
is screened, the regions of coherence
may be treated as independent one-dimensional
systems with an $XY$ degree of freedom.
The resulting effective Hamiltonian is
a sine-Gordon model, which supports
charge soliton excitations, whose energy
vanishes in the zero tunneling limit.
At zero temperature, the equivalent
1+1 dimensional system may be in a
state with either bound or unbound vortices.
Finally,
we argued that the former possibility
is an unusual situation which we call a
{\it ``critical Hall state''}, and is
characterized by a power
law tunneling $I-V$ characteristic.

We thank A.H.MacDonald, C.Tejedor, L.Mart\'{\i}n-Moreno and J.J.Palacios 
for useful discussions.
This work was supported by the CICyT of Spain under Contract No. PB96-0085,
and by the NSF under Grant No. DMR98-70681.

\newpage .
\vspace{-6.8truecm}
\begin{figure}
\epsfig{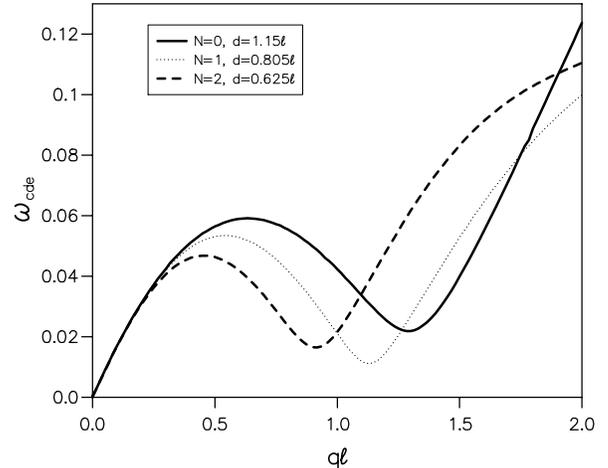}
\caption{
Dispersion of the charge density excitations as a function of the
wavevector $q$ for a double quantum well at filling factor 
$\nu = 4N+1$, and $t=0$.
The value of the layer separation is chosen to be near the critical
distance where the charge density instability occurs.
}
\end{figure}

\vspace{-4.8truecm}
\begin{figure}
\epsfig{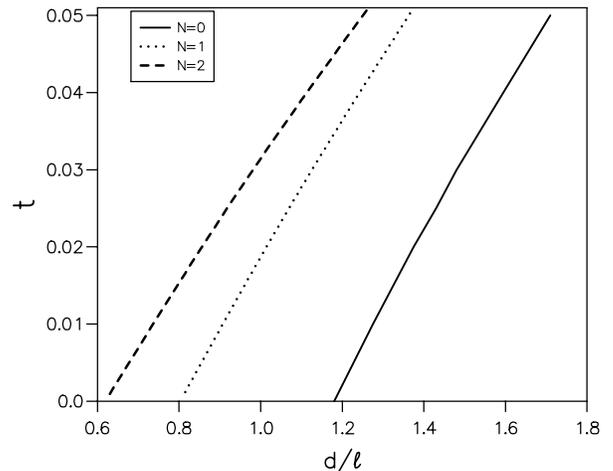}
\caption{
Phase diagram for the $\nu =4N+1$ state in double quantum well system. The
line separates the region where the  uniform coherent state occurs (small d and large $t$)
from the region where a translational broken symmetry ground state is expected (large $d$ and
small $t$). 
}
\end{figure}
\newpage .
\vspace{-6.8truecm}
\begin{figure}
\epsfig{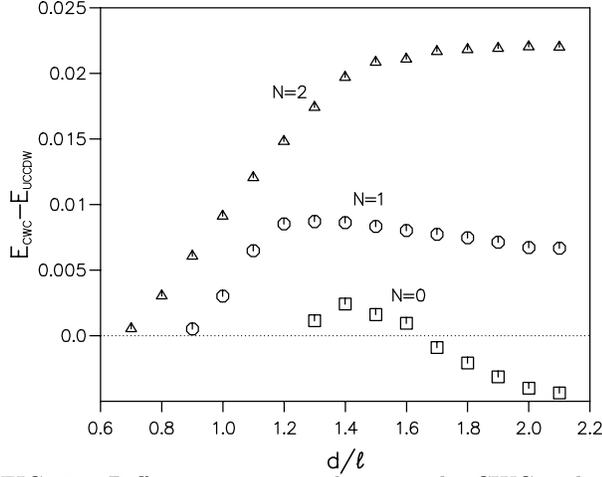}
\caption{
Difference in energy between the CWC and the UCCDW states as a function of $d$.
A positive(negative) value of this quantity means that the UCCDW state has lower 
(higher) energy than the CWC solution.}
\end{figure}
\vspace{-4.8truecm}
\begin{figure}
\epsfig{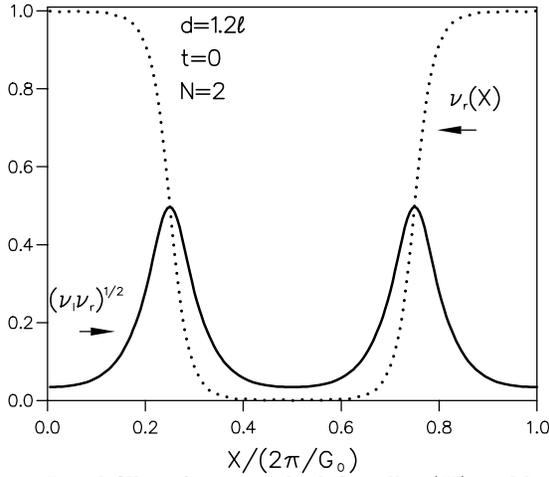}
\caption{
Local filling factor in the
left well $\nu _l (X)$ and local coherence
$\sqrt{ \nu _l (X) \nu _r (X)}$ as a function of $X$. 
These results correspond to $N=2$, $t=0$ and $d=1.2\ell$.
The length of the unit cell is $2 \pi /G_0$, and for this case
$G_0 \ell \sim 1$.
}
\end{figure}
\newpage .
\vspace{-6.8truecm}
\begin{figure}
\epsfig{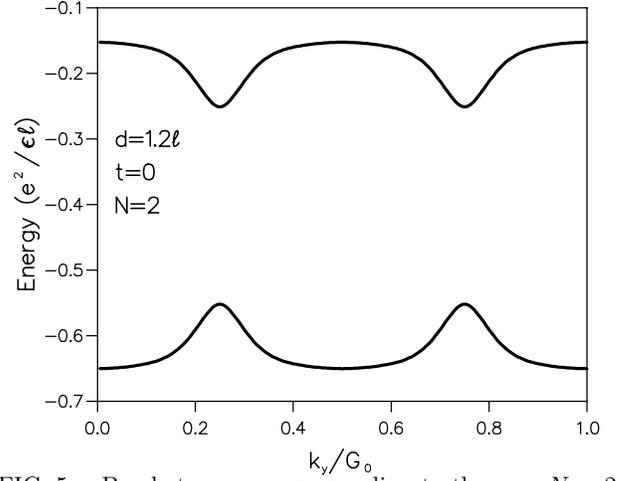}
\caption{ Band strucure corresponding to the case $N=2$, $d=1.2 \ell$ and $t=0$.
The eigenvalues depend on the momentum alon the $y$ direction, $X/\ell ^2$, which
is defined in the first Brilluin zone.
}
\end{figure}
\vspace{-4.8truecm}
\begin{figure}
\epsfig{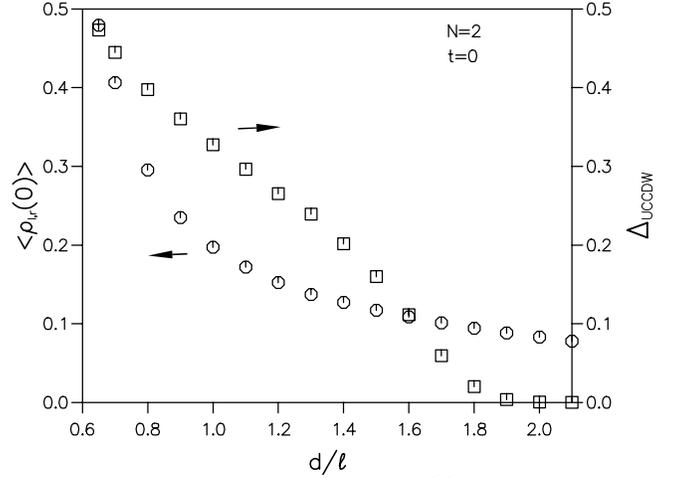}
\caption{ Interwell order parameter $<\rho _{l,r}(0)>$ and energy gap $\Delta _{UCCDW}$
as a function of the layer separation.
These results corresponds to the case $N=2$ and $t=0$. 
}
\end{figure}
\newpage .
\vspace{-6.8truecm}
\begin{figure}
\epsfig{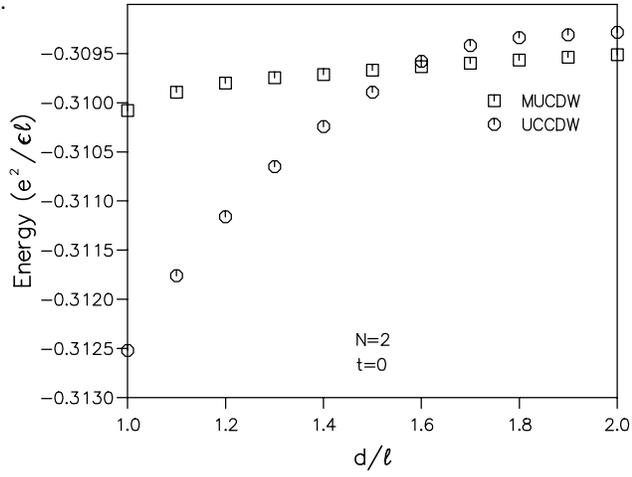}
\caption{
Energy as a function of the layer separation, for the MUCDW and UCCDW 
phase. The results corresponds to $N=2$ and $t=0$.
 }
\end{figure}
\vspace{-4.8truecm}
\begin{figure}
\epsfig{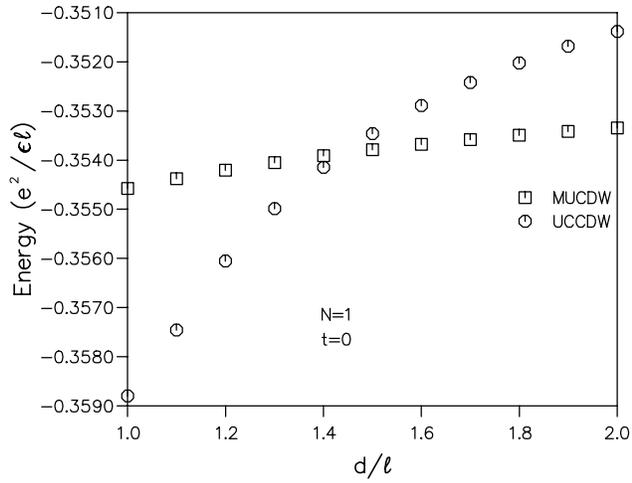}
\caption{ 
Energy as a function of the layer separation, for the MUCDW and UCCDW 
phase. The results corresponds to $N=1$ and $t=0$.
}
\end{figure}


\begin{references}
\bibitem{tilt} Anisotropic behavior may also be induced for
$N = 1$ in tilted samples.  See Ref. \cite{lilly2}.
\bibitem{com} In many situations, the regions will be coupled
by the long-range Coulomb interaction.  For the purposes of
this discussion, we ignore this coupling.  This may be accomplished
by the introduction of a screening plane parallel
to the layers.
\bibitem{com2} When Coulomb interactions between the linear coherent
regions are included, the coefficient $A$ is a function of $q_{\perp}$,
the wavevector describing the relative phases of the Goldstone
excitations in each region.
\end{references}
\end{document}